# Simplified Quantum Logic with Trapped Ions

C. Monroe, D. Leibfried, B.E. King, D.M. Meekhof, W.M. Itano, and D.J. Wineland

*National Institute of Standards and Technology, Boulder, CO  80303*

We describe a simplified scheme for quantum logic with a collection of laser-cooled trapped atomic ions. Building on the scheme of Cirac and Zoller, we show how the fundamental controlled-NOT gate between a collective mode of ion motion and the internal states of a single ion can be reduced to a single laser pulse, and the need for a third auxiliary internal electronic state can be eliminated.

Lately, much thought has been focussed on the implementation of simple quantum logic circuits for quantum computing [1] and other applications such as the generation of multi-particle entangled states for spectroscopy [2]. A promising candidate for quantum logic is based on the work of Cirac and Zoller [3], who showed how to construct universal multi-bit quantum logic gates in a system of laser-cooled trapped atomic ions. In the simplest form of the ion trap quantum computer, two internal electronic levels of each ion in a collection represent a quantum bit of information, and the quantum bits are "wired" together by virtue of the ions' collective motion in the trap. Trapped ions are attractive for quantum logic applications because their internal levels can be well isolated from the pernicious effects of quantum decoherence [4], while at the same time the ions strongly interact with one another through their Coulomb repulsion, allowing the formation of entangled states.

Several groups have shown that any quantum computation can be built from a series of one-bit and two-bit quantum logic gates [5]. A fundamental entangling quantum logic gate is the controlled-NOT (CN) gate [6,7], in which one quantum bit is flipped (rotated by $\pi$ radians) depending on the state of a second quantum bit. Cirac and Zoller showed how to realize a CN quantum logic gate in a collection of trapped ions by applying several appropriately tuned laser pulses to the ions and invoking an interaction with a third (auxiliary) internal atomic level. Their scheme was adapted to an experiment on a single $^9Be^+$ ion and required three laser pulses [8]. This report discusses a simpler CN gate scheme between an ion's internal and motional states which requires only a single laser pulse, while eliminating the requirement of the auxiliary internal electronic level. These simplifications are important for several reasons:

(1) Several popular alkali-like ion candidates, including $^{24}Mg^+$, $^{40}Ca^+$, $^{88}Sr^+$, $^{138}Ba^+$, $^{174}Yb^+$, and $^{198}Hg^+$, do not have a third electronic ground state available for the auxiliary level. These ions have zero nuclear spin with only two Zeeman ground states ($m_S = \pm\frac{1}{2}$). Although excited optical metastable states may be suitable for auxiliary levels in some of these ion species, this places extreme requirements for the frequency stability of the exciting optical field to preserve coherence.

(2) The elimination of an auxiliary ground state removes the possible existence of spectator internal atomic levels, which can act as potential "leaks" from the two levels spanned by the quantum bits (assuming negligible population in excited electronic metastable states). This feature may be important to the success of quantum error-correction schemes [9], which can be degraded when leaks to spectator states are present [10].

(3) The elimination of the auxiliary level can dramatically reduce the sensitivity of a CN quantum logic gate to external magnetic fields fluctuations. It is generally impossible to find three atomic ground states whose splittings are all first order magnetic field insensitive. However, for ions possessing hyperfine structure, the transition frequency between two levels can be made first order magnetic field insensitive at particular values of an applied magnetic field.

(4) Finally, a reduction of laser pulses simplifies the tuning procedure and may increase the speed of the gate. For example, the gate realized in Ref. [8] required the accurate setting of the phase and frequency of three laser pulses, and the duration of the gate was dominated by the transit through the auxiliary level.

For quantum logic with trapped ions, we assume that each ion has two internal states (denoted by $|\downarrow\rangle$ and $|\uparrow\rangle$ with energy separation $\hbar\omega_0$). We consider the center-of-mass (COM) mode of the ions' collective motion at harmonic frequency $\omega$, described by the ladder of states $|n\rangle$ having vibrational quantum number n = 0,1,2,... and energy $\hbar\omega(n+½)$. In quantum logic applications, we consider only the $|n\rangle=|0\rangle$ and $|n\rangle=|1\rangle$ motional states. The "reduced" CN logic gate flips the internal state of a particular ion j if and only if $|n\rangle=|1\rangle$, summarized in the following truth table and realized on a single ion in Ref. [8]:

$$\begin{aligned}|0\rangle|\downarrow\rangle_j &\rightarrow |0\rangle|\downarrow\rangle_j \\ |0\rangle|\uparrow\rangle_j &\rightarrow |0\rangle|\uparrow\rangle_j \\ |1\rangle|\downarrow\rangle_j &\rightarrow |1\rangle|\uparrow\rangle_j \\ |1\rangle|\uparrow\rangle_j &\rightarrow |1\rangle|\downarrow\rangle_j.\end{aligned} \quad (1)$$

When this gate is surrounded by two extra operations which map and reset the internal state of another ion i onto the collective COM state of motion, the result is a more general CN gate which flips the internal state of ion j if and only if ion i is in state $|\uparrow\rangle_i$ [3]. Below, we show how the reduced CN gate of Eq. (1) can be condensed to a single laser pulse, or, equivalently, how the more general CN gate between two ions can be condensed to three laser pulses.

If $|\downarrow\rangle_j$ and $|\uparrow\rangle_j$ are coupled by a dipole moment operator $\boldsymbol{\mu_j}$, then exposing ion j to a traveling-wave electric field $\mathbf{E}(z) = \mathbf{E_0}\cos(\mathbf{k}\cdot\mathbf{z}-\omega_L t+\phi)$ with frequency $\omega_L$, phase $\phi$, and wavevector $\mathbf{k}$, results in the interaction Hamiltonian:

$$\mathcal{H}_I^{(j)} = -\boldsymbol{\mu}_j\cdot\mathbf{E}(z) = -\hbar g_j(S_+^{(j)} + S_-^{(j)})(e^{i(\mathbf{k}\cdot\mathbf{z} - \omega_L t - \phi)} + e^{-i(\mathbf{k}\cdot\mathbf{z} - \omega_L t - \phi)}) \,. \quad (2)$$

In this expression, $g_j = \boldsymbol{\mu}_j\cdot\mathbf{E_0}/2\hbar$ is the resonant Rabi frequency connecting $|\downarrow\rangle_j$ to $|\uparrow\rangle_j$ in the absence of confinement, $S_+^{(j)}$ ($S_-^{(j)}$) is the internal level raising (lowering) operator of the jth ion, $\mathbf{z} = z_0\hat{\mathbf{z}}(a + a^\dagger)$ is the COM coordinate operator of the confined motion with associated harmonic raising (lowering) operator $a^\dagger$ (a) and zero-point spread $z_0 \equiv (\hbar/2M\omega)^{1/2}$, and M is the total mass of the ion collection. If the applied radiation frequency is tuned to $\omega_L = \omega_0 + (n'- n)\omega$, thereby coupling the states $|n\rangle|\downarrow\rangle$ and $|n'\rangle|\uparrow\rangle$, $\mathcal{H}_I$ is transformed to

$$\mathcal{H}_I^{(j)} = -\hbar g_j\left(S_+^{(j)} e^{i\eta(a+a^\dagger)-i\phi} + S_-^{(j)} e^{-i\eta(a+a^\dagger)+i\phi}\right) \quad (3)$$

in a frame rotating at $\omega_L$, where terms oscillating faster than $g_j$ ($g_j<<\omega,\omega_0$) have been neglected. Here, $\eta \equiv (\mathbf{k}\cdot\hat{\mathbf{z}})z_0$ is the Lamb-Dicke parameter, which controls the amount of coupling between internal and motional states. In the case of two-photon stimulated Raman transitions through a third (virtual) electronic level [11,12], $g_j$ is replaced by $g_{j,1}g_{j,2}/\Delta$, where $g_{j,1}$ and $g_{j,2}$ are the individual Rabi frequencies of the two beams when resonantly coupled to the virtual level and $\Delta$ is the detuning from the virtual level; $\omega_L$ ($\phi$) is replaced by the difference frequency (phase) of the beams; and $\mathbf{k}$ is replaced by the difference in wavevectors of the two beams $\delta\mathbf{k}=\mathbf{k_1}-\mathbf{k_2}$.

The CN quantum logic gate (Eq. (1)) can be realized with a single pulse tuned to $\omega_L = \omega_0$ (the "carrier") which couples the states $|n\rangle|\downarrow\rangle$ and $|n\rangle|\uparrow\rangle$ with Rabi frequency [13,14]

$$\Omega_{n,n} = \frac{1}{\hbar}|\langle n|\langle\uparrow|\mathcal{H}_I^{(j)}|\downarrow\rangle|n\rangle| = g_j|\langle n| e^{i\eta(a+a^\dagger)} |n\rangle| = g_j e^{-\eta^2/2} L_n(\eta^2) \,, \quad (4)$$

where $L_n(x)$ is a Laguerre polynomial. Specializing to the $|n\rangle=|0\rangle$ and $|n\rangle=|1\rangle$ vibrational levels relevant to quantum logic, we have

$$\begin{aligned}\Omega_{0,0} &= g_j e^{-\eta^2/2}, \\ \Omega_{1,1} &= g_j e^{-\eta^2/2}(1-\eta^2).\end{aligned} \tag{5}$$

From Eqs. (4) and (5), the carrier Rabi frequencies depend nonlinearly on the vibrational quantum number n, with the nonlinearity mediated by the Lamb-Dicke parameter η [15, 16]. The reduced CN gate (Eq. (1)) can be achieved in a single pulse by setting η so that $\Omega_{1,1}/\Omega_{0,0} = (2k+1)/2m$, with k and m positive integers satisfying k<m [17]. By driving the carrier transition for a duration τ such that $\Omega_{1,1}\tau = (k+½)\pi$, or a "π-pulse" (mod 2π) on the $|n\rangle=|1\rangle$ component, then $\Omega_{0,0}\tau = m\pi$. Thus the states $|1\rangle|\downarrow\rangle \leftrightarrow |1\rangle|\uparrow\rangle$ are swapped, while the states $|0\rangle|\downarrow\rangle$ and $|0\rangle|\uparrow\rangle$ remain unaffected. The net unitary transformation, in the {0↓, 0↑, 1↓, 1↑} basis is

$$\begin{bmatrix} 1 & 0 & 0 & 0 \\ 0 & 1 & 0 & 0 \\ 0 & 0 & 0 & ie^{i\phi}(-1)^{k-m} \\ 0 & 0 & ie^{-i\phi}(-1)^{k-m} & 0 \end{bmatrix}. \tag{6}$$

This transformation is equivalent to the reduced CN of Eq. (1), apart from phase factors which can be eliminated by the appropriate phase settings of subsequent logic operations [7].

The "magic" values of the Lamb-Dicke parameter which allow a single-pulse reduced CN gate satisfy $L_1(\eta^2) = 1-\eta^2 = (2k+1)/2m$ and are tabulated in Table I for the first few values [18]. It may be desirable for the reduced CN gate to employ the $|n\rangle=|2\rangle$ or $|n\rangle=|3\rangle$ state instead of the $|n\rangle=|1\rangle$ state for error-correction of motional state decoherence [19]. In these cases, the "magic" Lamb-Dicke parameters satisfy $L_2(\eta^2) = 1-2\eta^2+\eta^4/2 = (2k+1)/2m$ for quantum logic with $|n\rangle=|0\rangle$ and $|2\rangle$, or $L_3(\eta^2) = 1-3\eta^2+3\eta^4/2-\eta^6/6 = (2k+1)/2m$ for quantum logic with $|n\rangle=|0\rangle$ and $|3\rangle$.

Finally, we comment on the stability requirements of the Lamb-Dicke parameter in this scheme. Since the mapping pulses surrounding the single-ion CN gate (resulting in the general two-ion CN gate) have interaction strength proportional to η, this scheme does not place any additional premium on the stability of η. In this simplified scheme however, the *accuracy* of η must also be maintained; otherwise at least one of the two vibrational states will not be rotated by the correct amount. In the two-photon Raman configuration [11], the Lamb-Dicke parameter $\eta = |\delta \mathbf{k}| z_0$ can be controlled by both the frequency of the trap (appearing in $z_0$) and by the geometrical wavevector difference $\delta \mathbf{k}$ of the two Raman beams. Accurate setting of the Lamb-Dicke parameter should therefore not be difficult [16].

We acknowledge support from the U.S. National Security Agency, Office of Naval Research, and Army Research Office. We thank Brana Jelenković, Travis Mitchell, and Matt Young for critical reading of the manuscript.

**TABLE I**: Selected "magic" values of the Lamb-Dicke parameter η which satisfy $1-\eta^2 = (2k+1)/2m$. When the trapped ion is exposed to the carrier for a particular duration, the result is a "π-pulse" (mod 2π) between the $|1\rangle|\downarrow\rangle \leftrightarrow |1\rangle|\uparrow\rangle$ states and no net rotation (mod 2π) between the $|0\rangle|\downarrow\rangle \leftrightarrow |0\rangle|\uparrow\rangle$ states.

| k (rotation of $\|n\rangle=\|1\rangle$ states) | m (rotation of $\|n\rangle=\|0\rangle$ states) | $\eta = [1-(2k+1)/2m]^{1/2}$ |
|---|---|---|
| 0 (π) | 1 (2π) | 0.707 |
|  | 2 (4π) | 0.866 |
|  | 3 (6π) | 0.913 |
| 1 (3π) | 2 (4π) | 0.500 |
|  | 3 (6π) | 0.707 |
|  | 4 (8π) | 0.791 |
| 2 (5π) | 3 (6π) | 0.408 |
|  | 4 (8π) | 0.612 |
|  | 5 (10π) | 0.707 |
| 3 (7π) | 4 (8π) | 0.353 |
|  | 5 (10 π) | 0.548 |
|  | 6 (12π) | 0.645 |
| 4 (9π) | 5 (10π) | 0.316 |
|  | 6 (12π) | 0.500 |
|  | 7 (14π) | 0.597 |